\documentclass[aps,prb,twocolumn,amsmath,showpacs]{revtex4-1}
\usepackage{amsmath}
\usepackage{amssymb}
\usepackage{bm}
\usepackage{epsfig}
\usepackage{graphicx}
\usepackage{color}

\begin{document}

\title{Resources of polarimetric sensitivity in spin noise spectroscopy}

\author{P. Glasenapp$^1$, A. Greilich$^1$, I.~I. Ryzhov$^2$, V.~S. Zapasskii$^2$, \\
D.~R. Yakovlev$^{1,3}$, G.~G. Kozlov$^2$, and M. Bayer$^1$}

\affiliation{$^1$ Experimentelle Physik 2, Technische Universit\"at Dortmund, D-44221
Dortmund, Germany}

\affiliation{$^2$ St. Petersburg State University, Spin Optics
Laboratory, 198504 St. Petersburg, Russia}

\affiliation{$^3$ Ioffe Physical-Technical Institute, Russian
Academy of Sciences, 194021 St. Petersburg, Russia }

\begin{abstract}
We attract attention to the fact  that the ultimate
(shot-noise-limited) polarimetric sensitivity can be enhanced by
orders of magnitude leaving the photon flux incident onto the
photodetector on the same low level. This opportunity is of crucial
importance for present-day spin noise spectroscopy, where a direct
increase of sensitivity by increasing the probe beam power is
strongly restricted by the admissible input power of the broadband
photodetectors. The gain in sensitivity is achieved by replacing the
45$^{\circ}$ polarization geometry commonly used in
conventional schemes with balanced detectors by geometries with
stronger polarization extinction. The efficiency of these high-extinction
polarization geometries with enhancement of the detected signal by
more than an order of magnitude is demonstrated by measurements of
the spin noise spectra of bulk $n$:GaAs in the spectral range
835 -- 918\,nm. It is shown that the inevitable growth of the probe beam
power with the sensitivity gain makes spin noise spectroscopy
much more perturbative, but, at the same time, opens up fresh
opportunities for studying nonlinear interactions of strong light
fields with spin ensembles.
\end{abstract}


\maketitle

\section{Introduction}

The Faraday-rotation-based spin noise spectroscopy (SNS) first
proposed in 1981~\cite{AZ} survives nowadays its second birth. For
the last several years, due to a number of technical advancements,
the performance of SNS has been crucially improved in terms of both
its sensitivity and its
bandwidth~\cite{Romer,Muller:2010a,Crooker:2010}. The potential of
this novel technique has been convincingly demonstrated in numerous
experimental investigations (see, e.g.,
review~[\onlinecite{Muller:2010b}]). The most impressive
achievements were related to application of fast Fourier transform
(FFT) spectrum analyzers, which made it possible to shorten the
noise signal accumulation time by more than two orders of magnitude.
As a result, the SNS technique becomes more and more popular as an
efficient experimental tool for studying magnetic resonance and spin
dynamics in atomic and solid-state systems~\cite{OSN}.

At the same time, this technique, based on detecting random
statistical fluctuations of the spin-system implies measuring of
extremely small angles of Faraday or Kerr rotation and, therefore,
requires the highest polarimetric sensitivity. The polarimetric
sensitivity of optical measurements, defined by the smallest
detectable angle of polarization plane rotation, is known to be
fundamentally limited by the photon noise of the laser source or,
eventually, by the shot noise of the detector photocurrent. This
shot-noise limited polarimetric sensitivity in the visible spectral
range, being around $10^{-8} - 10^{-9}$\,rad, was achieved
in the late seventies of the last century~\cite{Aleks:1976,Jones}
and now serves as starting point of any experiment on SNS. The
main resource of increasing the shot-noise-limited polarimetric
sensitivity is related to the possible increase of probe beam power
controlling the shot noise level. However, the broadband
photodetectors used in the SNS (with a bandwidth of several hundreds
of MHz) usually have small photosensitive areas and cannot
endure light powers exceeding a few milliwatt. Thus, the
possibility of increasing the sensitivity in the most straightforward
way, by increasing the probe beam power in the scheme with a
standard polarization beamsplitter, appears to be strongly limited.
The question is whether one can enhance the polarimetric sensitivity
leaving the photocurrent of the detector and, therefore, its shot
noise on the same level. We show here that this task can be
efficiently solved using schemes with a high level of polarization
extinction.

Note that a similar experimental approach has been developed
earlier~\cite{Zap:1979,Zap:1982} and was mainly intended for
suppression of the excess intensity noise in polarimetric
measurements. We attract special attention to potentialities of
high-extinction polarimetry because it allows one to increase the
sensitivity by a few orders of magnitude through simplest, purely
polarization-based, means and surprisingly, has remained up to date
unexploited.

\section{Basic considerations}\label{sec:2}

Let us remind, first of all, how polarimetric sensitivity varies
with the angle $\varphi$ between the polarization plane of light
and the polarizing direction of the analyzer that converts polarization
oscillations of the light into oscillations of its intensity. The
standard polarization scheme employed in SNS (and, actually, in
most high-precision laser polarimeters) uses a polarizing
beamsplitter that combines two mutually orthogonal analyzers aligned
at $\pm 45^{\circ}$ with respect to the polarization plane of the
incident beam, see Fig.~\ref{fig:2}(a). The excess intensity noise
of the laser source is then suppressed by subtracting the photosignals
of the two outputs of the beamsplitter in a balanced detector. In
this scheme, the polarization-to-intensity converters operate in the
region of greatest steepness of the Malus law, which seems to be
most favorable from the viewpoint of polarimetric sensitivity. This
is, however, not exactly the case.

Consider an idealized situation when the polarizers are perfect (the
Malus law is valid with unlimited accuracy), and the shot noise of
the laser light is the only source of noise in the measuring system.
Let us write the Malus law in the form
\begin{equation}
N = N_0 \sin^2\varphi,
\label{eq:1}
\end{equation}
where $N_0$ and $N$ are the intensities of the
incident and transmitted light, respectively, expressed in the number of photons
(or photoelectrons) per second and $\varphi$ is the azimuthal angle of the
polarizing direction of the analyzer counted from the position of
total polarization extinction (``crossed'' position). Then, the
intensity response to a small rotation of the polarization plane
$\Delta\varphi$ (i.e., amplitude of the signal $A_s$) is given
by
\begin{equation}
A_s = \Delta N = 2 N_0 \sin\varphi \cos\varphi \Delta\varphi,
\label{eq:2}
\end{equation}
while the shot noise at a given angle $\varphi$ will be proportional
to the square root of the transmitted light intensity
\begin{equation}
A_n \equiv\sqrt{\langle(\delta N)^2\rangle} = \sqrt N = \sqrt{N_0}
\sin\varphi. \label{eq:3}
\end{equation}
So, for the signal-to-noise ratio, we have
\begin{equation}
A_s/A_n = 2 \sqrt{N_0}\cos\varphi \Delta\varphi.
\label{eq:4}
\end{equation}

Thus, we see that, in this idealized model, the polarimetric
sensitivity is the greatest for crossed position (at $\varphi  = 0$)
and remains practically the same at small $\varphi$ (it varies as $1
- \varphi^2/2$). In particular, at $\varphi = 45^{\circ}$, the
sensitivity decreases only by a factor of $\sqrt{2}$ compared with
that at $\varphi = 0$. This means, in turn, that in the framework of
this simplified model, the restrictions imposed on the input light
power of the photodetectors do not limit the polarimetric sensitivity of
the setup: the light power on the detector can always be decreased
with no loss of sensitivity (which is controlled only through $N_0$) by
decreasing the angle $\varphi$. Moreover, when passing from the
45$^{\circ}$ geometry to stronger polarization extinction by
decreasing the angle $\varphi$, we can considerably increase
intensity of the probe beam ($N_0$), leaving the light power on the
detector ($N$) the same.

Of course, in reality, the situation is different. First of all,
polarizers are not ideal. Their polarizing properties are
characterized by some finite extinction ratio $\zeta$ ($\zeta$, by
definition, is the ratio $N/N_0$ at $\varphi = 0$), usually lying in
the range of $10^{-4}-10^{-5}$. So, at small $\varphi$, the Malus
law is violated, and the above reasoning becomes inadequate for
$\varphi ^2 \lesssim \zeta$. Still, in many cases it is quite
realistic to increase the probe beam intensity by a few orders of
magnitude (leaving the photocurrent the same) and thus to increase
the polarimetric sensitivity by one or two orders of magnitude in
terms of the angles of Faraday rotation and by 2 - 4 orders of
magnitude in terms of the Faraday rotation noise power, provided
that the sample under study can withstand the appropriate light
power density.

\begin{figure}[t!]
\includegraphics[width=\columnwidth]{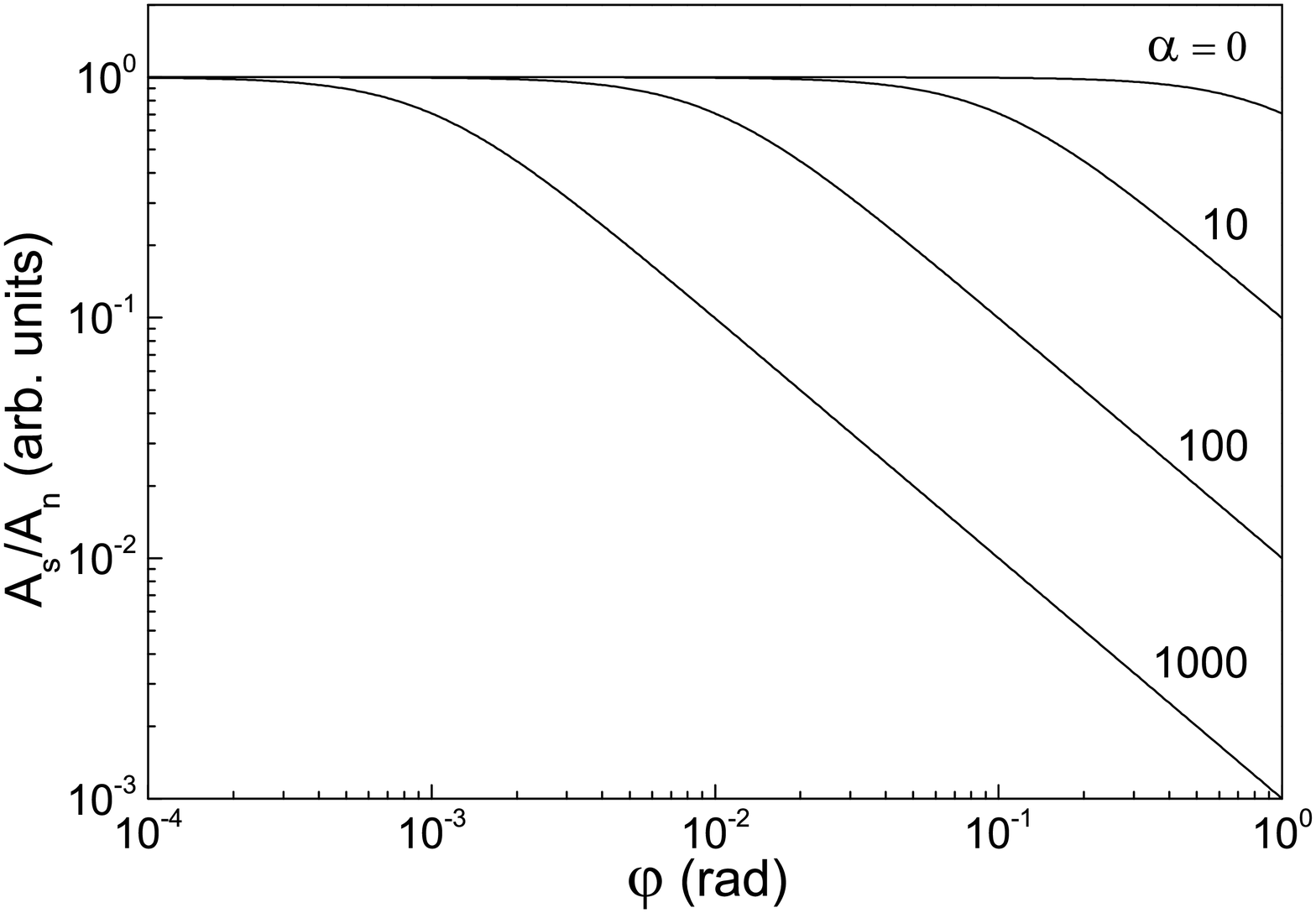}
\caption{Signal-to-noise ratio as function of $\varphi$ for
different excess noise values, $\alpha$. The effect of excess noise
is reduced for higher extinctions, $\varphi \rightarrow 0$.}
\label{fig:1}
\end{figure}

There is one more circumstance that distorts this idealized picture,
but, at the same time, provides some additional advantages for the
experimentalist. When the laser light exhibits excess intensity
fluctuations (which is often the case), the above treatment needs to
be corrected.  Now, the harmful intensity noise will contain a
statistical sum of the shot and the excess noise. It should be taken
into account that the excess noise, unlike the shot noise, varies
linearly with the light intensity, rather than in a square-root way.
So, in the presence of excess intensity noise, we have
\begin{eqnarray}\label{eq:5}
A_s\over A_n&=&\frac{2 N_0
\sin\varphi\cos\varphi\Delta\varphi}{\sqrt{N_0\sin^2\varphi+
{\alpha}^2N_0\sin^4\varphi}}\approx \nonumber \\
&\approx& \frac{2\sqrt{N_0}\Delta\varphi}{\sqrt{1+\alpha^2\varphi^2}}.
\end{eqnarray}
Here, $\alpha$ is the factor characterizing the amplitude of the excess
intensity noise of the incident beam \textit{in units of the shot-noise amplitude}.
Figure~\ref{fig:1} shows the dependence of $A_s/A_n$ on the
angle $\varphi$ for several values of the quantity $\alpha$.

It is important that due to different dependencies of the excess and
the shot noise on light intensity (and, therefore, on $\varphi$),
there exists an angle $\varphi$ for which the two contributions to
the noise become comparable, and then, at smaller $\varphi$, the
role of the excess noise becomes negligible. As seen from
Eq.~(\ref{eq:5}), this points are at $\varphi\alpha\sim 1$, when the
light intensity is attenuated through polarization crossing by a factor
of $\alpha^2$. In other words, by crossing the analyzer with the
light to be analyzed, we can, to a certain extent, suppress the
excess noise of the laser source and thus increase the polarimetric
sensitivity.


Summarizing the aforesaid, we conclude that by moving to the
geometry of high polarization extinction (HPE) we achieve three
goals: suppress the excess intensity noise of the laser source,
reduce the input light power on the photodetector down to an
acceptable level, and improve the sensitivity due to increased photon
flux through the sample. Indeed, by increasing the light power of
the probe beam by a factor of $k$ and leaving the input power on the
detector the same, we can increase the amplitude of the polarimetric
signal by a factor of $\sqrt{k}$ and its power by a factor of $k$.
It means that it is possible to increase the detected signal of the
noise power in the experiments on SNS by 2 - 3 orders of magnitude.
In most cases this looks realistic because the nonperturbative
detection of spin noise is usually performed in the region of
transparency, where the absorption of the sample is negligibly small and
the sample can be exposed to a high laser fluence.

\section{Polarimetric schemes}

There are several ways how to realize a HPE geometry. The most
straightforward implies detection of the light intensity behind
the analyzer fixed at an appropriate small angle $\varphi$ (nearly
``crossed'' geometry). A drawback of this simple scheme is that
after the analyzer all the information about motion of the
polarization plane of the incident beam appears to be converted to
variations of the light intensity (and the photocurrent), and all
resources of suppression of the excess intensity noise appear to be
exhausted. There exist, however, polarization schemes that make
it possible to additionally suppress the excess noise in the HPE
geometry.

\begin{figure}[t!]
\includegraphics[width=\columnwidth]{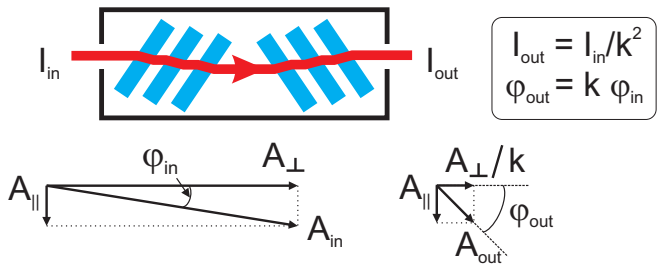}
\caption{ Effect of polarization pile on the polarization plane
rotation. When the polarization plane of the incident light is deviated
from the plane of extinction of the pile by an angle $\varphi$,
then, at the exit of the pile, this angle will be magnified by a
factor $k$ equal to the attenuation of the field component
$A_{\perp}$ along the extinction direction of the pile. Due
to the simultaneous decrease of the total light intensity by a factor of
$k^2$, the angle $k\cdot \varphi$ after the pile can be detected (in
the absence of excess noise) with the same signal-to-noise ratio but
at a lower level of light intensity.} \label{pile}
\end{figure}

One of them is the so-called polarization pile
method~\cite{Zap:1979,Zap:1982,pile}, when the polarization
attenuation of the light beam is achieved by transmitting it through
a properly aligned polarization pile, see Fig.~\ref{pile}. In this
case, the detected small angle of the polarization plane rotation
appears to be magnified by a factor equal to the square root of
intensity attenuation. This operation, in accordance with the
aforesaid, allows one to suppress the excess intensity noise, but
retains the possibility to suppress it further with the aid of a
balanced detector. This opportunity is important for noisy
laser sources, because the polarization extinction, in reality, can
suppress the excess noise only by 1 - 2 orders of magnitude, as
opposed to 4 orders of magnitude of noise suppression accessible for
a balanced detector~\cite{Aleks:1976,Arg}.

\begin{figure}[t!]
\includegraphics[width=\columnwidth]{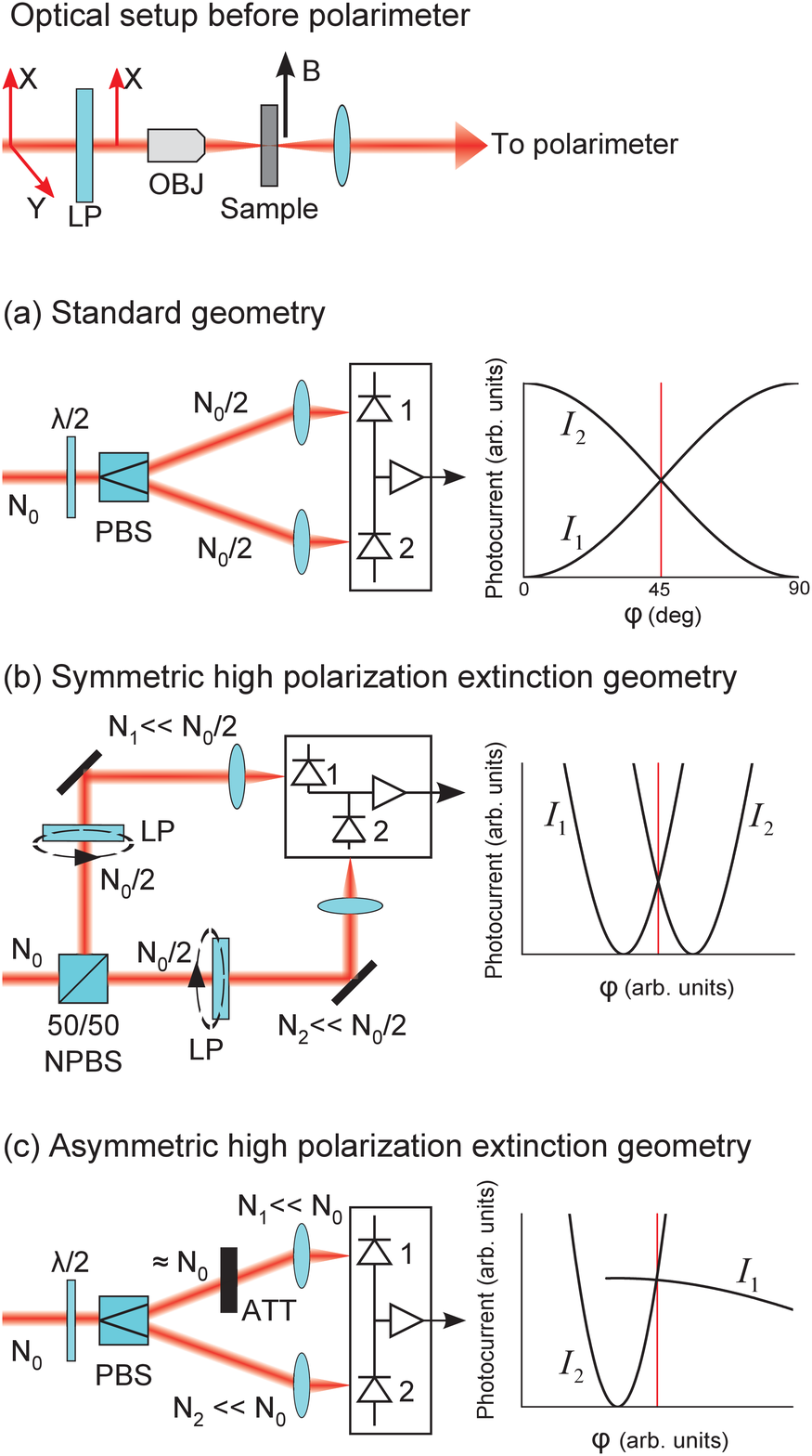}
\caption{Polarimetric schemes. Shown on top is the optical section
of the setup before the polarimetric detection in conventional spin
noise spectrometers (LP -- linear polarizer, OBJ -- microscope
objective). (a) Standard 45$^{\circ}$ polarimetric scheme with both
detectors operating at greatest steepness of the Malus law
($\lambda/2$ -- half-wave plate, PBS -- polarizing beamsplitter, e.
g., Wollaston prism). (b) Symmetric HPE scheme with nonpolarizing
beamsplitter (NPBS). The high polarization extinction is achieved by
adjusting the polarizers to the desired nearly crossed positions. In
this case, the dependence of the photocurrent disbalance on $\varphi$
(for the same photocurrent as in scheme (a)) becomes much steeper.
(c) Asymmetric scheme with a polarizing beamsplitter. The strong
disbalance in two arms of the PBS is compensated by the attenuating
filter ATT. The plots on the right show the photocurrents
$I_1$ and $I_2$ of the two detectors versus $\varphi$ in the appropriate
schemes.} \label{fig:2}
\end{figure}

There are other polarization schemes that combine polarization
extinction with excess noise subtraction. Figure~\ref{fig:2} shows
two possible arrangements of this kind, (b) and (c), together with
the standard 45$^{\circ}$ scheme in Fig.~\ref{fig:2}(a).

For the symmetric case of the HPE geometry shown in
Fig.~\ref{fig:2}(b), the analyzed laser beam is first split by a
polarization-insensitive beamsplitter into two beams of equal
intensity (50/50). Then both components are attenuated by polarizers
and coupled to inputs of the balanced detector. This scheme,
proposed earlier for suppressing the excess intensity noise (see
Ref.~[\onlinecite{Zap:1982}]), actually reproduces the standard
scheme (a) with stronger polarization extinction in each of the two
arms.

The other possible arrangement, Fig.~\ref{fig:2}(c), the asymmetric
case of the HPE geometry, can be easily obtained from the standard
45$^{\circ}$ scheme (Fig.~\ref{fig:2}(a)), by setting the
polarization beamsplitter into a strongly disbalanced position by
using the $\lambda/2$-plate. The intensity in the arm with low
polarization extinction is then attenuated by any optical filter to
achieve the balance. This scheme is highly attractive due to its
simplicity, because it can be easily obtained from the standard
45$^{\circ}$ geometry practically without changes in the setup. From
the viewpoint of sensitivity, these two geometries ((b) symmetric
and (c) asymmetric) are practically equivalent. More precisely, as
can be shown by simple calculations, the signal-to-noise ratios for
the two schemes are proportional to the factors:
\begin{eqnarray}
(A_s/A_n)_{sym}&=& 2\sqrt{N_0}\cos\varphi, \label{eq:b0}\\
(A_s/A_n)_{asym}&=& \sqrt{2N_0}/\cos\varphi, \label{eq:0}
\end{eqnarray}
respectively, and differ at small $\varphi$ (when the high-extinction
polarimetry makes sense) by a factor of $\sqrt{2}$, in favor of the
symmetric case.

In what follows, we demonstrate experimentally the efficiency of the HPE
schemes and discuss the increasing perturbativeness of SNS with
increasing light power density on the sample at high levels of
polarization extinction. Since the main part of the measurements was
performed using the ``symmetric'' scheme (Fig.~\ref{fig:2}(b)), we
will call this geometry simply HPE, unless specified otherwise.

\section{Experimental}

The experiments were performed with a well-studied bulk sample of
$n$:GaAs (sample B in Ref.~[\onlinecite{CrookerPRBbulk}]). The
antireflection-coated single crystal, 170\,$\mu$m in thickness, with
an electron density of $3.7\times 10^{16}$ cm$^{-3}$ was mounted on
the cold finger of a continuous flow helium cryostat at a
temperature of 5\,K. A linearly polarized light beam of a
frequency-stabilized Ti:sapphire continuous-wave ring laser with the
wavelength tuned below the GaAs bandgap ($\lambda > 820$\,nm) was
focused onto the sample with a microscope objective (upper sketch in
Fig.~\ref{fig:2}). The diameter of the light spot on the sample was
estimated to be about 10\,$\mu$m. The light beam transmitted through
the sample (see Fig.~\ref{fig:2}) passed through the beamsplitter
(either polarizing or polarization-insensitive, depending on the
type of the polarimetric scheme), and, after appropriate
polarization treatment, was detected by a 650\,MHz balanced
photoreceiver (New Focus 1607). The output signal of the photoreceiver
was amplified by 20\,dB and sent through a 580\,MHz low pass filter
(to avoid any undersampling) to a fast digitizer (2\,GS/s) with a
1\,GHz FFT processing unit implemented on a field programmable gate
array (FPGA). The system made it possible to process and average
the Fourier spectrum of the noise power density in real time (for more
details see Ref.~[\onlinecite{Crooker:2010}]).

The Faraday rotation noise spectra were measured in a transverse
magnetic field of 34\,mT, so that the Larmor frequency of the
electron spins ($g$-factor $|g_e|=0.415$), corresponding to the
central position of the spin noise resonance, lied in the range of
$\nu =$ 200\,MHz. To extract the signal we interleaved the applied
fields between 34 and 130\,mT, so that the spin noise resonance at
the higher field was shifted beyond the bandwidth of the detector,
leaving the background noise unaffected. The calibrated signal
($S(\nu)$) was obtained by dividing the spectrum with the noise
resonance at 200\,MHz ($P(\nu)$) by the one with the resonance
shifted to higher frequencies ($P_0(\nu)$):
\begin{equation}
S(\nu) = \left(\frac{P(\nu)}{P_0(\nu)}-1\right)\times 100\%
\label{eq:SNR}
\end{equation}
Since the main contribution to the background noise power, for our
conditions, comes from the shot noise, the signal calculated in this
way directly gives the spin noise in units of the shot noise power.

If not otherwise stated, the light power at the input of the
balanced detector was 1.5\,mW per channel. The accumulation times in the
measurements were varied between 1 and 3 minutes depending on the
signal amplitude.

\section{Results and discussion}
\subsection{Standard geometry}

Measurements of spin noise in the classical geometry
(Fig.~\ref{fig:2}(a)) performed in the spectral range 835 - 865\,nm
were aimed at obtaining reference spectra for comparison with the
results of application of the high-extinction polarization schemes.
\begin{figure}[t!]
\includegraphics[width=\columnwidth]{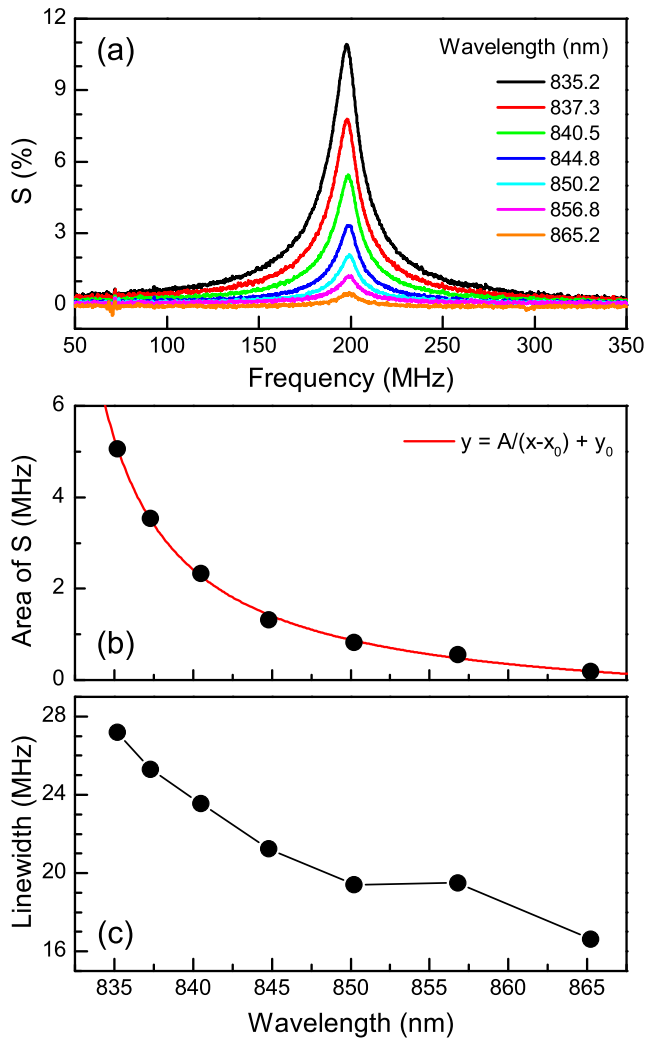}
\caption{Evolution of the SN spectrum , $S$, of $n$:GaAs with the
probe beam wavelength (a), spectral dependences of the SN area (b),
and SN linewidth (c) at low level of the probe beam power density
(6\,kW/cm$^2$) before the sample in the standard 45$^{\circ}$
geometry.}\label{WO}
\end{figure}

Figure~\ref{WO} shows the evolution of the spin noise (SN) power
spectrum ($S(\nu)$, see Eq.~(\ref{eq:SNR})) with the light
wavelength (a), the wavelength dependences of the integrated signal
(b), and the spin resonance linewidth (full width at half maximum)
(c). As expected~\cite{CrookerPRBbulk}, the amplitude of the SN
power rapidly decreases with increasing wavelength of the probe beam
(approximately inversely proportional to the detuning from the band
edge, see Fig.~\ref{WO}(b)). Starting with about 11\,\% of the spin
noise power amplitude at 835\,nm, we ended up with only 0.6\,\% at
865\,nm, where the resonance peak could still be reliably detected
with the used accumulation times. The only resource of sensitivity
remaining in this polarimetric scheme is related to increasing the
integration time $T_{int}$, which may improve the signal-to-noise
ratio in accordance with its square-root dependence on $T_{int}$.

The linewidth was extracted by fitting the curves in
Fig.~\ref{WO}(a) with a single Lorentzian. The increase in the
linewidth from 16.6\,MHz at 865\,nm to 27.2\,MHz at 835\,nm
qualitatively coincides with the known experimental
data~\cite{CrookerPRBbulk} and thus reflects the perturbative effect of
the probe beam. At the same time, the absolute values of the widths
appear to be larger than reported in
Ref.~[\onlinecite{CrookerPRBbulk}]. We ascribe this difference to
the smaller size of the spot created by the microscopic objective and,
correspondingly, to a higher power density, which, in our experiment
was about 6\,kW/cm$^2$ as compared with 1\,kW/cm$^2$ in
Ref.~[\onlinecite{CrookerPRBbulk}]. In addition, the light spot
diameter of the laser beam was comparable with the diffusion length
of electrons in this sample, which should be on the order of
10\,$\mu$m~[\onlinecite{DiffusionLength}] and may additionally
contribute to the line broadening.

\subsection{High polarization extinction geometries}

\begin{figure}[t!]
\includegraphics[width=\columnwidth]{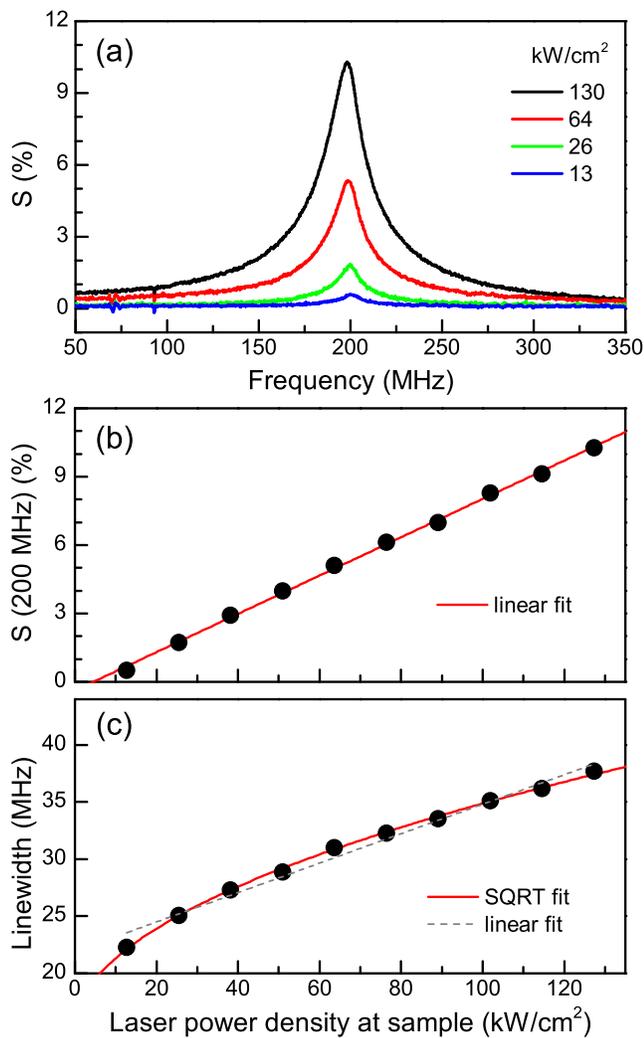}
\caption{Effects of light power density on the spin noise spectra
measured in the HPE scheme at $\lambda=865$\,nm. (a) Evolution of
the SN spectra $S$. (b) Amplitude of the signal $S(\nu)$ at
$\nu=200$\,MHz versus light power density together with a linear
fit. (c) Power dependence of the linewidth fitted by a square root
function. Dashed line is a linear fit shown for comparison.}
\label{CPO}
\end{figure}

The HPE schemes imply increasing intensity of the laser probe with
simultaneous polarization attenuation of the input light on the
photodetector. So, for our first demonstration of efficiency of this
technique in detecting SN spectra of $n$:GaAs, we have chosen a
wavelength of 865\,nm, at which the SN resonance measured in
standard geometry, at a laser power density of 6\,kW/cm$^2$ was
below 1\% of shot noise. Figure~\ref{CPO} shows the results of the
measurements in the range of laser power density up to
130\,kW/cm$^2$ (corresponding to 100\,mW for 10\,$\mu$m spot),
measured before the sample.

One can see a drastic enhancement of the signal, from $\sim$1\,$\%$
to $\sim$10\,$\%$, with increasing laser power by a factor of 10,
Figs.~\ref{CPO}(a) and~\ref{CPO}(b). It is important, that this
growth of the signal exactly corresponds to the growth of the
signal-to-noise ratio, because the shot-noise level is controlled by
the noise of the detector photocurrent, which, at a fixed incident
power of 1.5\,mW per channel, remains the same. It should be noted,
that these measurements were made in the range of angles $\varphi$
not smaller than 0.2\,rad, i.e., still far from the limit imposed by
the performance of the polarizers. In other words, the resources of
polarimetric sensitivity, in these experiments, are far from being
exhausted. The linear dependence of the SN signal on the laser power
density in Fig.~\ref{CPO}(b) fully corresponds to the description
presented in Sec.~\ref{sec:2}.

Of course, one should bear in mind that the gain in sensitivity does not come for free.
The increasing polarization extinction is needed
to attenuate the growing photon flux that probes the sample and may
strongly perturb it. Indeed, the increase of the probe beam power
increases the perturbative broadening of the spin resonance, noticeable
even at lower intensities in this spectral range in the standard
geometry (Fig.~\ref{WO}(c)). With increasing power density, the spin
resonance linewidth varies approximately in a square-root way from
22.3\,MHz at 13\,kW/cm$^2$ to 37\,MHz  at 130\,kW/cm$^2$, see
Fig.~\ref{CPO}(c). This broadening noticed in
Ref.~[\onlinecite{CrookerPRBbulk}] is presumably related to residual
absorption of the $n$:GaAs crystal, which remains essential even at
large detuning. We also suspect that the square-root dependence of
the spin resonance linewidth detected in the SN spectrum is not
incidental and may reflect a linear dependence on the light field
amplitude (or on the Rabi frequency), characteristic of certain
processes of resonant interaction of a monochromatic wave with
inhomogeneously broadened ensembles~\cite{Khodovoi}. This
assumption can be confirmed or rejected only after additional
experimental and theoretical studies.

\begin{figure}[t!]
\includegraphics[width=\columnwidth]{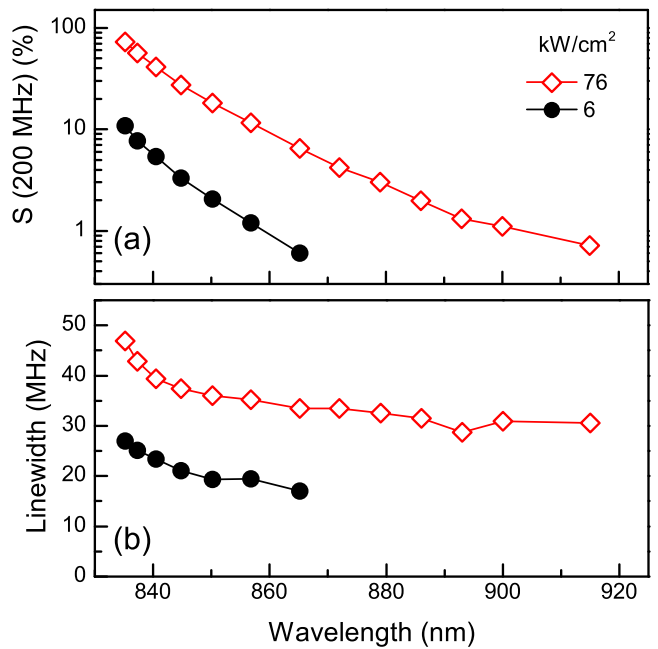}
\caption{Comparison of the signal amplitude (a) and linewidth (b)
for the standard geometry with power density of 6\,kW/cm$^2$ and the
HPE geometry with power density of 76\,kW/cm$^2$ on the sample. The
enhancement of the signal allows one to realize a much larger
detuning of the laser from the band edge. Lines are guide to the
eye.} \label{WAVDEP}
\end{figure}

Additional information about light-induced broadening of the
spin-noise spectra and their behavior at large detuning was obtained
from spectral measurements performed in a much wider range of
wavelengths, accessible due to enhanced sensitivity. The
measurements were performed in the HPE geometry at a laser power
density of 76\,kW/cm$^2$. Starting at 835\,nm, to have a direct
comparison with the standard 45$^{\circ}$ scheme at the power
density of 6\,kW/cm$^2$, it was possible to extend the wavelength
range up to 918\,nm, still having $0.7\,\%$ of signal amplitude (in
units of the shot noise power). The results of these
measurements, shown in Fig.~\ref{WAVDEP}(a), demonstrate, as we
believe, in a spectacular way the potential of the new
experimental approach.

Under conditions of a fixed input power on the detector, the
increase of the detected spin noise power in the high-extinction
geometry, as was already mentioned, exactly corresponds to the
increase in the signal-to-noise ratio. Due to increasing absorption
at wavelengths shorter than 845\,nm, the polarization
extinction, needed to maintain the same laser power on the detector,
slightly decreases, thus decreasing the gain in sensitivity. This,
however, did not distort noticeably the results of the measurements
in the short-wavelength region where the signal amplitude, $S(\nu)$,
measured in this geometry exceeded $\sim70$\,$\%$.

Figure~\ref{WAVDEP}(b) shows the dependence of the linewidth versus
wavelength over a wide spectral range. Along with the total increase of
broadening at higher laser power densities, we see, in a more
impressive form, the characteristic spectral dependence of the
broadening effect. A remarkable feature of this dependence is that
it becomes practically wavelength-independent at longer wavelengths.
It also correlates with the assumption that the light-induced
broadening of the SN spectra is related to the effects of
non-resonant perturbation of the spin-system through optical
absorption, rather than to resonant effects, which should become
sensitive to detuning.

\begin{figure}[t!]
\includegraphics[width=\columnwidth]{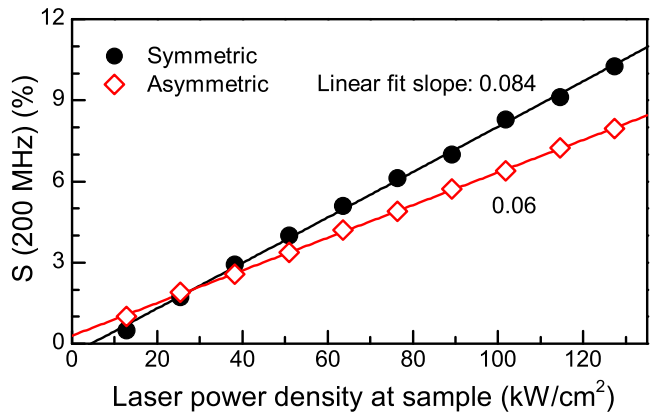}
\caption{Variation of the SN signal with angle $\varphi$ for the two HPE
geometries: symmetric and asymmetric. $\lambda=865$\,nm.}
\label{CPBW}
\end{figure}

For completeness of the methodological analysis of polarimetric
sensitivity, we performed measurements in both HPE geometries
considered above -- symmetric (Fig.~\ref{fig:2}(b)) and asymmetric
(Fig.~\ref{fig:2}(c)). The results of the measurements shown in
Fig.~\ref{CPBW} completely agree with our previous conclusions. As
mentioned above (see Eqs.~(\ref{eq:b0}) and~(\ref{eq:0})), at small
$\varphi$, the signal-to-noise ratio in the symmetric scheme appears
to be higher by a factor of $\sqrt 2$ than in the asymmetric scheme.
As the laser power is decreased, the angle $\varphi$ is increased to
keep the power on the diodes constant. This corresponds to a lower
level of polarization extinction, where the approximation of small
$\varphi$ is violated, and starting from $\varphi = 0.57$\,rad (when
$\cos^2\varphi = 1/\sqrt 2$) the asymmetric geometry becomes more
preferable. Indeed, in the vicinity of this angle, which corresponds
to a probe power density of 30\,kW/cm$^2$, the two experimental
dependencies, approximated by straight lines, intersect. At these
angles, however, the HPE geometry, which implies strong polarization
attenuation of the light intensity, looses its sense.

\subsection{Thermal effects}

\begin{figure}[t!]
\includegraphics[width=\columnwidth]{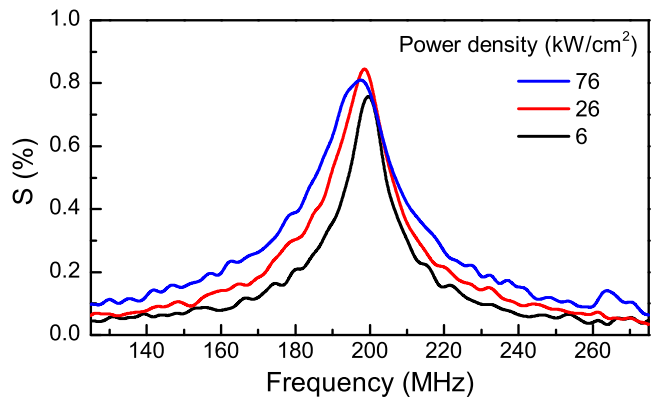}
\caption{Spin noise spectra (obtained in standard geometry) at
different light power densities on the sample with no polarization
enhancement. The input power on the detector was maintained at low levels
with an attenuating filter. $\lambda=865$\,nm.}\label{CGATT}
\end{figure}

To evaluate possible thermal effects in the apparent
enhancement of the SN signal (which we attributed entirely to
enhancement of polarimetric sensitivity) we performed a simple
experiment: in the measurements made in the standard 45$^{\circ}$
scheme, we varied the light intensity on the sample keeping the intensity
before the beamsplitter on the same level (3\,mW). In this case, we
have no polarization enhancement, and, in the absence of any optical
nonlinearity, all the spectra recorded in this way should be
identical. The results of these measurements presented in
Fig.~\ref{CGATT} show, however, an increase in the linewidth for
higher laser powers without any essential  changes in the amplitude of
the signal. It means, that even if the sample were heated by the
high laser power, this could not lead to the observed dramatic changes
in the signal magnitude, as shown in Fig.~\ref{CPO}(a).

\begin{figure}[t!]
\includegraphics[width=\columnwidth]{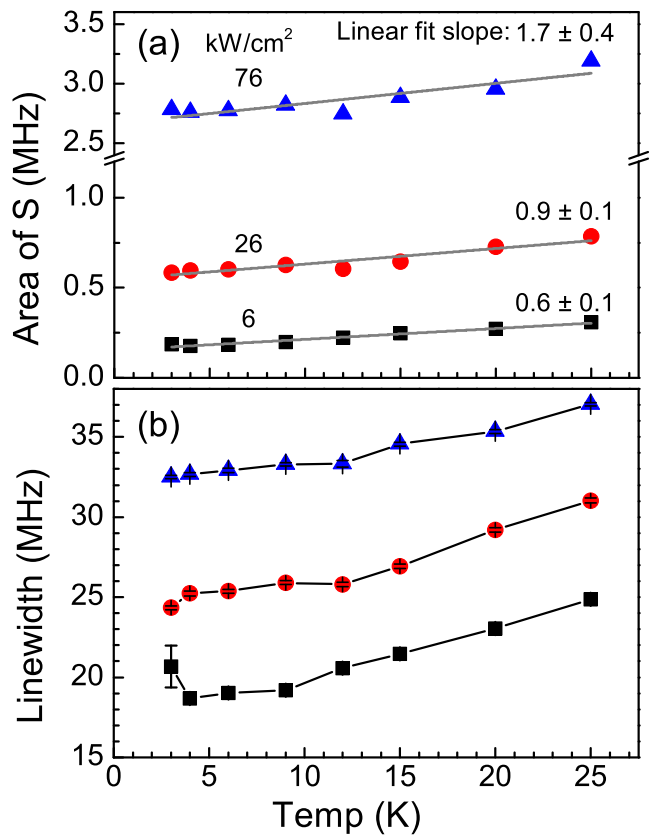}
\caption{Temperature dependence of SN area (a) and linewidth (b) at
different levels of the probe beam power density. Lines in (b) are
guide to the eye. $\lambda=865$\,nm.} \label{TEMP}
\end{figure}

Additional information about the role of thermal effects in formation of
the SN spectra was obtained from studies of the temperature dependence
in the range 3 -- 25\,K. The measurements were performed
with the laser beam power density on the sample varying from 6 to
76\,kW/cm$^2$. The results are presented in Fig.~\ref{TEMP}.

For our sample, it is known that only a fraction of the total number
of electrons within the thermal energy $k_B T$ around the Fermi
energy can fluctuate~\cite{CrookerPRBbulk}. As seen from
Fig.~\ref{TEMP}(a) the SN area for all power densities depends
linearly on temperature, having very similar slopes. It is therefore
reasonable to assume, that the increased laser power does not induce
local heating of the electron system, which would increase the
contributing range (i.e. the number) of fluctuating electrons, and
therefore reduce the effect of the external temperature increase.

At high laser powers, the SN area increases much more than any
temperature in the studied temperature range may provide, as one can
see from Fig.~\ref{TEMP}(a). One can also see from
Fig.~\ref{TEMP}(b) that the linewidth of 32 MHz observed at
76\,kW/cm$^2$ at 4\,K, could be achieved with 6\,kW/cm$^2$ at
around 50\,K only. So, it seems highly probable that the nature of the
light-induced perturbation of the SN spectrum in this case is not
related to effects of thermal heating, while the responsible
mechanism remains obscure.

\section{Conclusions}

In this paper we have shown how application of HPE geometries may
widen the possibilities of the Faraday-rotation-based spin noise
spectroscopy. The main merit of the HPE schemes arises from the
opportunity to increase the intensity of the probe laser beam and,
correspondingly, to increase the ultimate polarimetric sensitivity
controlled by the photon noise, without increasing the photon flux
incident on the photodetector. This additional degree of freedom,
which is absent in the commonly applied 45$^{\circ}$ polarization
geometry, affects simultaneously two characteristics of the
measurement system: it enhances the polarimetric sensitivity and increases
the level of perturbation of the spin-system by the probe beam. The
SNS technique, evidently, ceases to be ``nonperturbative'', but
acquires new qualities, which it initially was not intended for: It
turns into a tool for studying processes of nonlinear interaction of
strong optical fields with spin-systems. We demonstrate here this
possibility by experimental studies of SN spectra of $n$:GaAs.

We consider the proposed experimental approach to be highly
important for the further development of spin noise spectroscopy and
for broadening the scope of its application. Suffice it to say that
the gain in sensitivity provided by the HPE schemes may be as high
as that achieved with the advent of FFT spectrum analyzers, which
had brought a real breakthrough into the field of the spin noise
spectroscopy. Now, in combination with the most advanced methods of
data acquisition, SNS will be able to considerably widen the range
of objects for study and the scope of physical problems amenable to
this technique.

\subsection*{Acknowledgements}
The authors thank S.~A. Crooker for fruitful discussions and for
providing the sample. This research was supported by the Deutsche
Forschungsgemeinschaft and BMBF. The financial support from the
Russian Ministry of Education and Science (contract No.
11.G34.31.0067 with SPbSU and leading scientist A.~V. Kavokin) is
acknowledged. V.~S.~Z. acknowledges support through a Gambrinus
fellowship of TU Dortmund University.

\end{document}